\documentclass[%
	reprint,
superscriptaddress,
 amsmath,amssymb,
 aps,
 prl,
]{revtex4-1}

\usepackage{graphicx}
\usepackage{dcolumn}
\usepackage{bm}
\usepackage{units}
\usepackage[colorinlistoftodos]{todonotes}
\usepackage{textcomp}
\usepackage{hyperref}
\usepackage{xcolor}
\hypersetup{
    colorlinks,
    linkcolor={red!50!black},
    citecolor={blue!50!black},
    urlcolor={blue!80!black}
}

\renewcommand\d{\ensuremath{{\rm d}}}

\begin{document}


\title{Bright x-ray radiation from plasma bubbles in an evolving laser wakefield accelerator
}

\newcommand{\JAI}{The John Adams Institute for Accelerator Science, Imperial College London, London, SW7 2AZ, UK}

\newcommand{\GOLP}{GoLP/Instituto de Plasmas e Fus\~{a}o Nuclear, Instituto Superior T\'{e}cnico, U.L., Lisboa 1049-001, Portugal}

\newcommand{\CLF}{Central Laser Facility, STFC Rutherford Appleton Laboratory, Didcot OX11 0QX, UK}

\newcommand{\UCL}{Department of Physics and Astronomy, University College London, London WC1E 6BT, UK}

\newcommand{\LMU}{Fakult\"at f\"ur Physik, Ludwig-Maximilians-Universit\"at M\"unchen, D-85748 Garching, Germany}
\newcommand{\MPQ}{Max-Planck-Institut f\"ur Quantenoptik, Hans-Kopfermann-Str. 1, D-85748 Garching, Germany}

\newcommand{\DESY}{Deutsches Elektronen-Synchrotron DESY, Notkestr. 85, 22607 Hamburg, Germany}

%
%
%

\author{M.S.~Bloom}
\affiliation{\JAI}

\author{M.J.V.~Streeter}
\email{m.streeter09@imperial.ac.uk}
\affiliation{\JAI}

\author{S.~Kneip}
\affiliation{\JAI}

\author{R.A.~Bendoyro}
\affiliation{\GOLP}

\author{O.~Cheklov} 
\affiliation{\CLF}

\author{J.~M.~Cole}
\affiliation{\JAI}

\author{A.~D\"opp}
\affiliation{\JAI}
\affiliation{\LMU}
\affiliation{\MPQ}

\author{C.J.~Hooker}
\affiliation{\CLF}

\author{J.~Holloway}
\affiliation{\UCL}

\author{J.~Jiang}
\affiliation{\GOLP}

\author{N.C.~Lopes}
\affiliation{\JAI}
\affiliation{\GOLP}

\author{H.~Nakamura}
\affiliation{\JAI}

\author{P.A.~Norreys}
\affiliation{\CLF}

\author{P.P.~Rajeev}
\affiliation{\CLF}

\author{D.R.~Symes}
\affiliation{\CLF}

\author{J.~Schreiber}
\affiliation{\JAI}
\affiliation{\LMU}

\author{J.~C.~Wood}
\affiliation{\JAI}

\author{M.~Wing}
\affiliation{\UCL}

\author{Z.~Najmudin}
\affiliation{\JAI}

\author{S.P.D.~Mangles}
\email{stuart.mangles@imperial.ac.uk}
\affiliation{\JAI}

\date{\today}

\begin{abstract}
We show that the properties of the electron beam and bright x-rays produced by a laser wakefield accelerator can be predicted if the distance over which the laser self-focuses and compresses prior to self-injection is taken into account. 
A model based on oscillations of the beam inside a plasma bubble shows that performance is optimised when the plasma length is matched to the laser depletion length. 
With a 200~TW laser pulse this results in an x-ray beam with median photon energy of \unit[20]{keV}, $> 6\times 10^{8}$ photons above \unit[1]{keV} per shot and a peak brightness of $\unit[3 \times 10^{22}]{photons~s^{-1}mrad^{-2}mm^{-2} (0.1\% BW)^{-1}}$.

\end{abstract}

\maketitle

Laser wakefield accelerators~\cite{Tajima1979} have gathered increasing interest since it was first shown that they were capable of producing high quality electron beams~\cite{Mangles2004, Geddes2004, Faure2004}. 
Development has continued apace, and laser wakefield accelerators can now produce ultra-short bunches of electrons, down to a few femtoseconds~\cite{Lundh2011}, and reach multi-GeV beam energies~\cite{Leemans2014,Wang2013a, Kim2013EnhancementPulses}.
One of the primary near-term uses of laser wakefield accelerators is the production of bright, femtosecond duration pulses of broadband x-rays~\cite{Rousse2004, Kneip2010b}, that are suitable for a range of applications~\cite{Albert2014}. 

A laser wakefield accelerator is formed when an intense, short-duration laser pulse is fired into a moderate density plasma.
The ponderomotive force associated with the laser pushes plasma electrons out of its way as it propagates.
The much heavier positive ions are effectively stationary and so the electrons are pulled back towards their equilibrium positions once the laser has passed.
The resulting collective charge oscillation has a relativistic phase velocity in the wake of the laser. 
When driven by a sufficiently intense laser pulse, almost all of the electrons can be expelled from an approximately spherical cavity behind the drive, known as the plasma bubble~\cite{Pukhov2002}.
The electric fields inside this bubble are capable of accelerating electrons to \unit[$\sim1$]{GeV} in just \unit[$\sim1$]{cm}~\cite{Kneip2008}. 

Electrons can be self-injected into the bubble from the background plasma if the wave has a sufficiently high amplitude~\cite{Mangles2012}. 
The three-dimensional structure of the bubble means that the injected electron beam undergoes strongly non-linear betatron oscillations with a short wavelength (\unit[$\sim 1$]{mm} for \unit[$\sim1$]{GeV}) and so generates synchrotron x-rays in the multi-keV spectral range~\cite{Rousse2004}.

In this letter we report on the experimental optimisation of the x-rays generated by a laser wakefield accelerator driven by a 200~TW laser.
We show that the electron and x-ray properties are well described by an analytical model that includes the fact that self-injection only occurs after the pulse has self-focused and self-compressed to a sufficiently high intensity.
We also show that this source outperforms previous reports of x-ray emission from wakefield accelerators~\cite{Kneip2010b, Wang2013a}.

The experiment was performed using the Astra Gemini laser at the Rutherford Appleton Laboratory, which delivered pulses of \unit[12]{J} and duration \unit[55]{fs} \textsc{fwhm} onto a supersonic gas jet target.
The laser was focused with an $f$/20 off-axis parabolic mirror to a  \unit[22]{\textmu m} \textsc{fwhm} spot containing $\alpha$ $\approx 30\%$ of the energy.
The peak intensity at focus in vacuum was $I \approx \unit[2.2 \times 10^{19}]{W cm^{-2}}$, corresponding to a normalised vector potential $a_{\rm 0} = eA_{\rm 0}/(m_{\rm e} c^2) \simeq 3.0$.

Two different helium gas jet targets with exit diameters of \unit[(10, 15)]{mm}  were used, producing approximately uniform density plasmas of length  \unit[(8.5, \,13)]{mm} with electron densities up to  \unit[$(8.0, \, 4.0) \times 10^{18}$]{cm$^{-3}$} respectively. 
The laser was focused onto the front edge of the gas flow. 
The generated electron beam was analysed using a magnetic spectrometer consisting of a \unit[30]{cm} long \unit[1]{T} permanent dipole magnet and two scintillating (Kodak Lanex regular) screens.

The magnet also swept the electron beam away from an indirect detection x-ray \textsc{ccd} camera (Princeton Instruments PIXIS) placed on the laser axis. 
This was mounted outside the vacuum chamber behind a \unit[180]{\textmu m} thick Be window. 
An array of 16 metallic filters mounted on a thin Al/mylar substrate was placed directly in front of the camera's CsI scintillator.
A schematic of the experimental layout is available here~\cite{SuppMat}.

The x-ray spectrum was found by performing a least squares fit to the signal detected behind each filter, taking into account the transmission through each filter and the absolutely calibrated detector response under the assumptions that the spectrum has a synchrotron-like shape given by:
$d^2I/(dEd\Omega)_{\theta = 0} \propto \xi^2 \mathcal{K}^2_{2/3}(\xi/2)$, where $\mathcal{K}_{2/3}(x)$ is a modified Bessel function of order $2/3$ and $\xi = E/E_{\rm c} $. 
The shape of this spectrum is characterized by a single parameter, the critical energy, $E_{\rm c}$  \footnote{This definition of $E_{\rm c}$ is different to that used in refs~\cite{Esarey2002a, Kneip2008} but consistent with refs~\cite{Rousse2004, Rousse2007, Jackson1999})}, which was assumed to be constant over the detector. 
Gaps between the x-ray filters and repeated filters allowed gradients in the x-ray beam profile to be taken into account.

\begin{figure}
\includegraphics[width= 8.64cm]{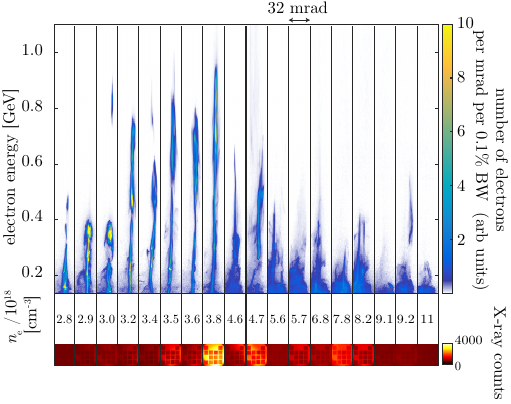} 
\begin{center}
\caption{Variation of electron and x-ray beams as a function of $n_{\rm e}$ for 10 mm nozzle. 
Top: Dispersion-corrected spectrally-dispersed images of the electron beam for a selection of shots. 
The horizontal axis represents the angle at which electrons exit the accelerator in the non-dispersion plane.
Bottom: x-ray CCD camera images for the same shots.}
\label{ebeam_montage}
\end{center}
\end{figure}

Fig.\,\ref{ebeam_montage} shows the variation of the electron beam spectrum and corresponding x-ray \textsc{ccd} images with plasma density for the \unit[10]{mm} nozzle. 
Shots in this data sub-set correspond to the brightest x-ray image recorded at each density. 
The data shows that there is an optimum density for acceleration of $n_{\rm e} \approx \unit[3.8 \times 10^{18}]{cm^{-3}}$, and that the x-ray signal is correlated with the electron beam energy.
Above the optimum density, the electron beam  begins to develop transverse structure and increased divergence. 
This is consistent with the electron beam interacting with the plasma, driving its own wake and becoming susceptible to propagation instabilities~\cite{Huntington2011}.
As the electron beam dephases, it can also interact with the laser field~\cite{Mangles2006a} which can also increase the x-ray flux~\cite{Cipiccia2011}. 
However, under these conditions no major enhancement in x-ray flux is evident due to these effects. 

\begin{figure}[htb]
\includegraphics[width= 8.64cm]{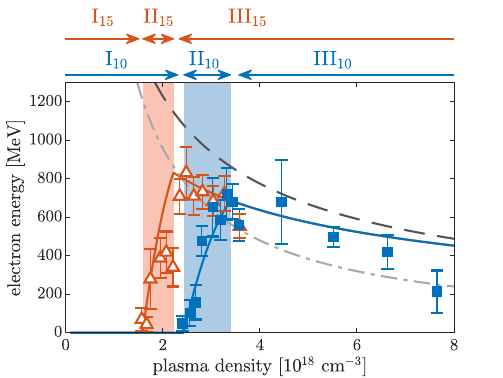} 
\begin{center}
\caption{Variation of $W_{\rm max}$ with $n_{\rm e}$ for 10 (squares) and 15 mm (triangles) gas jets.
Each point is the mean from $N$ = 1--13 shots, bars represent the combined statistical and measurement errors.
Curves show maximum energy predicted using: eqn.\,\ref{WeiLu} (dot-dash); eqn.\,\ref{with_sf} (dash); eqn.\,\ref{PIPE} for the \unit[10]{mm}  (blue); and \unit[15]{mm} (red) nozzles.
The regions (I, II and III) described in the text are indicated for each nozzle. 
Region II for each nozzle is also indicated as a shaded region.}
\label{ebeam_energies}
\end{center}
\end{figure}

Fig.\,\ref{ebeam_energies} shows the variation of the electron energy, $W_{\rm max}$ with $n_{\rm e}$ for both the 10 and \unit[15]{mm} gas nozzles.
Both datasets show a similar trend; above a threshold density, $W_{\rm max}$ rapidly increases until it reaches a maximum after which it decreases approximately $\propto 1/n_{\rm e}$.
Both the threshold and optimum densities occur at lower densities for the \unit[15]{mm} nozzle. 
As the x-ray signal is optimised when the electron beam energy is highest, optimisation of the x-ray generation relies on understanding how to optimise the electron energy. 
The maximum energy predicted by Lu \emph{et al.},~\cite{Lu2007}: 
\begin{align}
W_{\rm max}/{\left(m_{\rm e} c^2\right)} &= {\textstyle \frac{2}{3}}a_{\rm 0}({n_{\rm c}/{n_{\rm e}}}) \,, 
\label{WeiLu}
\end{align}
is plotted in fig.\,\ref{ebeam_energies} (dot-dash line) using the $a_{\rm 0}$ for vacuum.
Here $n_{\rm c} = \epsilon_{\rm 0} m_{\rm e} \omega^2 / e^2$ is the critical plasma density for radiation with angular frequency $\omega$.
Eqn.\,\ref{WeiLu} consistently underestimates the electron energy at high densities and does not predict an optimum density. 

Eqn.\,\ref{WeiLu} assumes a non-evolving laser pulse. 
However, self-focusing and pulse compression occur as the pulse propagates in its self-generated wakefield, causing $a_{\rm 0}$ to increase. 
The maximum value of $a_{\rm 0}$ can be calculated by assuming that the pulse evolves to a matched spot size that satisfies $a_{\rm 0} \approx 2 (P_{\rm f}/P_{\rm c})^{1/3}$~\cite{Lu2007}, where $P_{\rm c} = (2 m_{\rm e} c^3/r_{\rm e}) n_{\rm c} /n_{\rm e}$ is the critical power for relativistic self-focusing and $r_{\rm e}$ is the classical electron radius. 
The final power, $P_{\rm f}$, depends on pulse compression and the amount of laser energy that is trapped in the bubble. For a fraction $\alpha$ of the total laser energy $E_{\rm L}$ compressed to a pulse duration $\tau_{\rm f}$,  the resulting expression is:
\begin{align}
a_{\rm max} &\simeq 2 \left(\alpha E_{\rm L}/({\tau_{\rm f} P_{\rm c}) }\right)^{\frac{1}{3}} \,.
\end{align}
Even though this expression does not include the effects of pump depletion and photon deceleration, it has been shown to be sufficiently accurate to predict the self-injection threshold~\cite{Mangles2012}.
The pulse compression can be quantified by assuming that the front of the pulse travels at the linear group velocity in the plasma while the rear of the pulse, which sits in the significantly reduced plasma density inside the bubble, travels at $c$. 
For an initial pulse duration $\tau_{\rm 0}$, the pulse duration after propagation length $l$ is then: $\tau_{\rm f}(l) \approx \tau_{\rm 0} - {n_{\rm e} l}/({2 c n_{\rm c}})$~\cite{Schreiber2010}.
The maximum propagation length, $L_{\rm max}$, will be limited by pump depletion $L_{\rm pump} \approx c\tau_{\rm 0}\, n_{\rm c} / n_{\rm e}$~\cite{Lu2007} or by the length of the target, $L_{\rm target}$, if $L_{\rm target} < L_{\rm pump}$.
So accounting for pulse evolution, the beam energy varies as:
\begin{align}
W_{\rm max}^\prime/(m_{\rm e} c^2) &\approx {\textstyle \frac{4}{3}}\left(\alpha E_{\rm L}/(\tau_{\rm f} P_{\rm c}) \right)^\frac{1}{3} (n_{\rm c}/n_{\rm e})
\label{with_sf}
\end{align}

Eqn.\,\ref{with_sf} is shown as the dashed line in fig.\,\ref{ebeam_energies}.  
This model overestimates the observed energy gain, only approaching the data at high densities.
It also still fails to explain the initial increase in beam energy with increasing density. 
These features can be explained by including a distance over which the $a_{\rm 0}$ amplification occurs before self-injection. 
We call this the pre-injection pulse evolution length, $L_{\rm PIPE}$.  
This \textsc{pipe} length will decrease at higher densities as the pulse evolution rates increase~\cite{Mori1997b}.  
The variation of electron energy with density can be split into three regions. 
At low density (region I), the \textsc{pipe} length is longer than the gas jet and so no electrons are injected.

As the density is increased, the evolution becomes fast enough that injection occurs before the end of the gas jet, resulting in low energy electron beams. This region II is marked by the shaded area in fig.\,\ref{ebeam_energies}.
Increasing the density further reduces the  \textsc{pipe} length and brings the injection point earlier in the gas jet. But in region II, the density is low enough that the laser has not depleted by the end of the gas jet, so earlier injection leads to an increase in acceleration length. This coupled with the increase in the accelerating field ($\propto \sqrt{n_{\rm e}}$) results in a rapid increase of beam energy with increasing density.

Once the density is high enough that the pump depletion length is less than the gas jet length (region III), increasing the density actually decreases the length over which the electron beam is accelerated. 
Therefore in region III, despite the continuing increase in electric field strength, higher density results in a decrease in the electron energy. 
At the highest densities, the \textsc{pipe} length is very short and the electron beam energy approaches the dephasing limit (eqn.\,\ref{with_sf}). 
To model all this behaviour, we need an expression for the energy gained by an electron accelerated over some fraction of a dephasing length, and an expression that tells us what fraction of a dephasing length is available after the \textsc{pipe} length is taken into account.

In the bubble regime, the longitudinal electric field, which is responsible for acceleration, is  linearly proportional to the longitudinal distance from the bubble centre, i.e. $E_z(\xi)  \propto \xi$ , where $ \xi = z - v_\phi t$ and $v_\phi$ is the velocity of the bubble \cite{Kostyukov2004}.  
Because injected electrons travel faster than the bubble they experience a decreasing accelerating electric field as they advance relative to the bubble.  
The energy gained by the electron is found by integrating the accelerating force over the path of the electron.
To first order $\mathrm{d}\xi/\mathrm{d}z \approx (c-v_{\phi})/c$ and so the electron energy is quadratic with acceleration distance \cite{Thomas2010e}.
The maximum energy is reached when an electron travels from the back to the centre of the bubble, i.e. when 
$\Delta \xi = r_b/2$ which occurs when the acceleration length is equal to the dephasing length, $L_{\phi} = \frac{4}{3}a_{\rm 0}^{1/2} (n_{\rm c}/n_{\rm e}) c/\omega_{\rm p}$.
Because injection occurs after the \textsc{pipe} length, the fraction of the dephasing length available for acceleration is therefore
$\Delta_{\rm acc} = (L_{\rm max} - L_{\rm PIPE})/L_\phi$, 
where $L_{\rm max}$ is the maximum length available for the laser wakefield accelerator:  the shorter of $L_{\rm target}$ and $L_{\rm pump}$. 
The effect of the pre-injection evolution phase is therefore to reduce the maximum energy reached compared to Eqn.\,\ref{with_sf} according to
\begin{align}
W_{\rm max}'' = W_{\rm max}'\left( 2 \Delta_{\rm acc} - \Delta_{\rm acc}^{2}\right).
\label{PIPE}
\end{align}

Eqn.\,\ref{PIPE} is plotted in fig.\,\ref{ebeam_energies} (solid red and blue lines) with a single fitting parameter, \unit[$S = 11.8 \pm 1.8$]{\textmu m}, chosen to best reproduce the experimental trend using a least squares fit.
As self-focusing happens more quickly than pulse compression in this regime, the \textsc{pipe} length is closely related to the length over which pulse amplification occurs, $L_{\rm evol}$, which can be calculated using the model in ref~\cite{Streeter2018PRL}.
That predicts that \unit[$S = (n_e/n_c)L_{\rm evol} = 11$]{\textmu m} for \unit[$n_e = 2 \times 10^{18}$]{cm$^{-3}$} with a weak dependence on plasma density. 
The \textsc{pipe} model reproduces all of the main features of the experimentally observed variation in electron energy including: the rapid rise to an optimum density; the reduction in optimum density for the longer nozzle; and the slower fall-off of maximum energy at higher densities.
The model performs equally well for both the 10 and \unit[15]{mm} nozzles with the same value for $S$, indicating that the precise shape of the gas density plays a minor role in determining the evolution as compared to the initial laser parameters. 

Perhaps counter intuitively, this shows that for our value of $S$ the maximum electron energy from a fixed length target occurs at too low a density for acceleration over a full dephasing length, i.e.~$\Delta_{\rm{acc}} < 1$.
Rather the optimum occurs at the lowest density where the acceleration is limited by pump depletion, i.e.~when $L_{\rm{PIPE}} + \Delta_{\rm{acc}}L_\phi  = L_{\rm{target}} = L_{\rm{pump}}$.
In general for any value of $S$, the electron energy from a fixed length target is maximized for the density at which $L_{\rm{target}}=L_{\rm{pump}}$ or $ L_{\rm{target}} = L_{\phi} - L_{\mathrm{PIPE}}$, whichever occurs at a lower density.
Increasing the length of the target allows for higher electron energies at a lower density (as shown by the \unit[15]{mm} nozzle data), but this cannot be done indefinitely as at some density the pulse will never evolve to the point of injection.

\begin{figure}[ht!]
\begin{center}
\includegraphics[width=8.64cm]{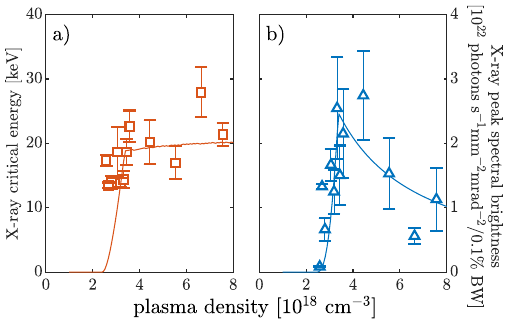}
\caption{
Observed x-ray: a) $E_{\rm c}$  and b) $\mathcal{B}_{\rm 0}$ as a function of $n_{\rm e}$ for the 10 mm nozzle. 
Each point is the mean of $N =$ 1--10 shots, bars represent the combined statistical and measurement errors.
Curves show predictions using eqns.\,\ref{betatronBrightness} \& \ref{spectral_shape}. 
}
\label{x-ray}
\end{center}
\end{figure}

Fig.\,\ref{x-ray} plots how the x-ray critical energy, $E_{\rm c}$, and peak brightness, $\mathcal{B}_{\rm 0}$, vary with $n_{\rm e}$ for the \unit[10]{mm} gas nozzle.
Both the critical energy and brightness show similar behaviour; rapidly increasing as the density is increased before turning over above $n_{\rm e} \approx \unit[3.8 \times 10^{18}]{cm^{-3}}$.
The critical energy reaches $E_{\rm c} \approx \unit[30]{keV}$.
The peak brightness, calculated assuming a constant duration matching that of the laser pulse (\unit[55]{fs}) and an r.m.s. source radius consistent with previous measurements (\unit[1]{\textmu m}), reaches $\unit[3 \times 10^{22}]{photons~s^{-1}mrad^{-2}mm^{-2} (0.1\% BW)^{-1}}$. 
This is significantly higher than previous results at lower laser power, e.g.~\cite{Kneip2010b}, primarily due to the ability to accelerate electrons to $\sim$~GeV energies, thus increasing their ability to radiate.
Under optimum conditions, a total photon yield of $N_{\rm X}  \approx 6 \times 10^{8}$ per shot  above \unit[1]{keV} was measured.  

By solving the equation of motion for an electron in the fields of a plasma bubble, Thomas~\cite{Thomas2010e} calculates the spectrum of radiation emitted by summing the synchrotron spectrum emitted at each bend in the trajectory as the electron is accelerated and then decelerated from the back of the bubble to the front, i.e.~over a distance $L = 2L_{\rm \phi}$.
The number of betatron oscillations is found to be, $N_\beta = \gamma_{\rm p}$, where $\gamma_{\rm p}$ is the Lorentz factor associated with the bubble motion. 
The resulting x-ray spectrum is synchrotron-like with an enhanced high-energy tail.
We modify the expression in \cite{Thomas2010e} to take into account that in general the acceleration does not take place over a full dephasing length.
The modified spectrum for a beam charge of $eN_{\rm b}$ is:
\begin{align}
\label{betatronBrightness}
 \frac{\d^2 I}{\d\omega \d\Omega} & = \gamma_{\rm p} \frac{3e^2}{\pi^3\epsilon_{\rm 0} c} N_{\rm b} \left(\frac{W_{\rm max}^\prime}{m_{\rm e} c^2}\right)^2 \mathcal{A} \left(\frac{\omega}{2\omega_{\rm c0} },\Delta_{\rm acc}\right) \,,
\end{align}
where $E_{\rm c0} = \hbar \omega_{\rm c0} = 7.75 \times 10^{-4} a_{\rm 0}^{5/2}(n_{\rm c}/n_{\rm e})^{9/8}$ is the critical energy of a synchrotron spectrum corresponding to that emitted by an electron at the maximum energy $W_{\rm max}^\prime$.
The spectral shape function, $\mathcal{A}(\xi, \Delta_{\rm acc} )$ is: 
\begin{align}
\label{}
 \mathcal{A} &= \xi^2\int_{-1}^{\Delta_{\rm acc} - 1} (1-x^2)^{-\frac{3}{2}} \mathcal{K}_{2/3}^2\left( \xi( 1 - x^2)^{-\frac{7}{4}}\right) \d x \,.
 \label{spectral_shape}
\end{align} 

The calculated variation in $E_{\rm c}$ and $\mathcal{B}_{\rm 0}$ have been overlaid on the experimental data in fig.\,\ref{x-ray}a \& b, for a plasma length of \unit[8.5]{mm} with $L_{\rm PIPE} = Sn_{\rm c}/n_{\rm e}$ and $S = $\unit[11.8]{\textmu m}.
The critical energy curve requires no additional fitting parameter beyond the \textsc{pipe} length scaling already determined from the electron data. 
In the absence of an absolute charge calibration, we have used the scaling law of Lu \emph{et al.}~\cite{Lu2007}  multiplied by a fitting parameter $\hat{N}$, $N_{\rm b}= 3.1 \times 10^{8} \hat{N} \lambda_{\rm 0} \sqrt{P_{\rm f}}$  in the  calculation of peak brightness.
Taking into account the effect of pulse compression, this gives an approximately constant charge of $\approx \unit[44]{pC}$ over the density range of interest.  
The peak brightness using this model is maximal at the density for which the pump depletion length equals the given target length.
The models for the electron energy, x-ray critical energy and density dependence of the x-ray brightness are all consistent with the experimental data using the single fitting parameter $S$.

\begin{figure}
\begin{center}
\includegraphics[width=8.64cm]{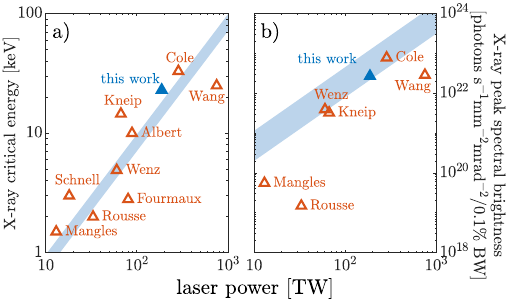}
\caption{Scaling of x-ray radiation with laser power from various experiments~\cite{Mangles2009, Rousse2004, Schnell2012a,Kneip2010b,Wenz2015,Fourmaux2011a,Albert2013a,Wang2013a,Cole2015}:  a) $E_{\rm c}$; b) $\mathcal{B}_{\rm 0}$. 
The shaded region corresponds to eqns.\,\ref{Ec_scaling} and \ref{B_scaling} for $n_{\rm th} \le n_{\rm e}  \le 3n_{\rm th}$.}
\label{scaling_figure}
\end{center}
\end{figure}

Fig.\,\ref{scaling_figure} compares our measured $E_{\rm c}$ and $\mathcal{B}_{\rm 0}$ with previous experiments.
Kneip et al.~\cite{Kneip2012b} calculated the following scaling laws in terms of the laser power, $P$ (in TW) for both $E_{\rm c}$ (in keV) and $\mathcal{B}_{\rm 0}$ (in $\unit[]{photons~s^{-1}mrad^{-2}mm^{-2} (0.1\% BW)^{-1}}$):
\begin{align}
E_{\rm c}  &= 3.5 \times 10^{-2} \left( \beta \alpha P\right)^\frac{5}{6} \left(n_{\rm c}/n_{\rm e}\right)^\frac{7}{24}\,,
\label{Ec_scaling} \\
 \mathcal{B}_{\rm 0}  &= 2.7 \times 10^{19} \hat{N}\left( \beta \alpha P\right)^\frac{2}{3}\left(n_{\rm c}/n_{\rm e}\right)^\frac{19}{12}\,,
 \label{B_scaling}
\end{align}
where a power amplification factor, $\beta$ is included to account for pulse compression. 
Compression  over a depletion length at the rate in~\cite{Schreiber2010} gives $\beta = 2$.

Eqns.\,\ref{Ec_scaling} and \ref{B_scaling} treat $P$ and $n_{\rm e}$ as independent parameters. 
However, increasing $P$ decreases the threshold density for self-injection, $n_{\rm th}$, so experiments at higher power typically operate at lower density~\cite{Mangles2016}. 
Using~\cite{Mangles2012} to find $n_{\rm th}(P)$, we can therefore eliminate $n_{\rm e}$ from the above expressions. 
Furthermore experiments are typically optimised just above the threshold, so the curves for $E_c$ and $\mathcal{B}_{\rm 0}$ (plotted in fig.\,\ref{scaling_figure}) span the range $[n_{\rm th} , 3 n_{\rm th}]$.

The scaling laws and experimental data both show that higher power lasers produce higher energy and brighter x-ray sources.  
However it is important to note that experiments can differ significantly from the scaling law.
For example, higher photon energies can be produced if asymmetries are present in the wake, which increases the betatron oscillation amplitude.
However, this also increases the source size, decreasing the brightness~\cite{Mangles2009}.  

The photon source in our experiments is much brighter, and at higher photon energy than those at lower power~\cite{Kneip2010b} and comparable to an unoptimised experiment at five times higher laser power~\cite{Wang2013a}.
Our experimental data and the model developed here suggest that a significant increase in the flux of x-ray radiation from a laser wakefield accelerator can be achieved by careful optimisation of the length and plasma density for a given laser system. This would greatly broaden the range of applications that are accessible with theses sources~\cite{Albert2014}.

The authors thank the staff at the Central Laser Facility, Rutherford Appleton Laboratory for their assistance. 
This research was supported by STFC (ST/J002062/1, ST/P002021/1),  EPSRC (EP/I014462/1), and the European Research Council (ERC) under the European Union's Horizon 2020 research and innovation programme (Grant Agreement Nos. 682399 and 653782).
The authors confirm that the all data used in this study are available without restriction.  
Data and analysis codes can be obtained using the following link \url{https://doi.org/10.5281/zenodo.3765331}.

\bibliography{Mendeley_Xray_PIPE_paper_edited.bib}

\end{document}